\documentclass[aps,prl,twocolumn,showpacs]{revtex4-1}
\usepackage{epsfig}
\usepackage{bm}
\usepackage{color}

\newcommand{\be}{\begin{eqnarray}}
\newcommand{\ee}{\end{eqnarray}}
\newcommand{\no}{\nonumber}
\newcommand{\diff}{\mathrm d}
\newcommand{\e}{\mathrm e}

\begin{document}
\title{Quantum-Classical Correspondence of Shortcuts to Adiabaticity}
\author{Manaka Okuyama}
\author{Kazutaka Takahashi}
\affiliation{Department of Physics, Tokyo Institute of Technology, 
Tokyo 152-8551, Japan}

\date{\today}

\begin{abstract}
We formulate the theory of shortcuts to adiabaticity in classical mechanics.
For a reference Hamiltonian, 
the counterdiabatic term is constructed from 
the dispersionless Korteweg--de Vries (KdV) hierarchy.
Then the adiabatic theorem holds exactly 
for an arbitrary choice of time-dependent parameters.
We use the Hamilton--Jacobi theory to define the generalized action.
The action is independent of the history of the parameters 
and is directly related to the adiabatic invariant.
The dispersionless KdV hierarchy 
is obtained from the classical limit of 
the KdV hierarchy for the quantum shortcuts to adiabaticity.
This correspondence suggests some relation between 
the quantum and classical adiabatic theorems.
\end{abstract}

\maketitle

Shortcuts to adiabaticity (STA) is a method controlling dynamical systems.
The implementation of the method results in 
dynamics that are free from nonadiabatic transitions 
for an arbitrary choice of time-dependent parameters 
in a reference Hamiltonian.
It was developed in quantum systems~\cite{DR1,DR2,Berry,CRSCGM}
and its applications have been studied 
in various fields of physics and engineering~\cite{STA}. 
It is important to notice that this method,
decomposing the Hamiltonian into the reference term and the counterdiabatic term, 
is applied to any dynamical systems
and offers a novel insight into the systems.

It is an interesting problem 
to find the corresponding method in classical mechanics 
from several points of view as we discuss in the following.
Jarzynski studied STA for the classical system 
by using the adiabatic invariant~\cite{Jarzynski}.
He found the form of the counterdiabatic term in a certain system 
based on a generator of the adiabatic transport.
Although several applications have been discussed~\cite{DJdC,PJ,JDPS}, 
the complete formulation of the classical STA is still under investigation.

In the quantum system, the adiabatic theorem is described
by the adiabatic state constructed from 
the instantaneous eigenstate of a reference Hamiltonian $H_0$.
When the time-dependence of the parameters in the Hamiltonian is weak,
the solution of the Schr\"odinger equation can be approximated 
by the adiabatic state.
On the other hand, the adiabatic theorem in the classical system
is described by the phase volume defined in periodic systems.
The closed trajectory in phase space for a fixed parameter 
gives the adiabatic invariant
\be
 J = \int \diff x\diff p\,\theta(E_0-H_0), \label{J}
\ee
where $E_0$ denotes the instantaneous energy.
$J$ is defined instantaneously and
the adiabatic theorem states 
that $J$ is approximately conserved 
when the parameter change is slow.

Thus the quantum and classical adiabatic theorems look very different and
the relation between them is not obvious.
The quantum STA is introduced so that the quantum adiabatic theorem 
holds exactly and we expect that the same holds for the classical case.
These two formulations will allow us to make a link between two theorems.
In this letter, we develop the theory of the classical STA.
First, we formulate the classical STA in a general way 
so that the counterdiabatic term can be 
calculated, in principle, from the derived formula.
Second, we show that the adiabatic theorem holds exactly in the classical STA 
and the adiabatic invariant is obtained directly from 
the nonperiodic trajectory.
Third, we show that the quantum STA reduces to the classical STA 
by taking the limit $\hbar\to 0$, which suggests some relation between 
quantum and classical adiabatic dynamics.


We consider classical systems with one degree of freedom for simplicity.
The system is characterized by the Hamiltonian 
$H =H(x,p;\alpha(t))=H_0+H_{\rm CD}$.
The dynamical variables in phase space are denoted by $x$ and $p$.
To consider the counterdiabatic driving, 
we use a time-dependent parameter $\alpha(t)$.
It is a straightforward task to generalize the present formulation to 
systems including several parameters.
In the adiabatic time evolution, 
$\alpha(t)$ represents a slowly-varying function.
Here, we do not impose any conditions on $\alpha(t)$.
In the quantum counterdiabatic driving, the state is determined
instantaneously.
The same must be implemented for classical dynamics and 
we impose the condition 
that, 
when the solution of the equation of motion 
$(x,p)=(x(t;\alpha(t)),p(t;\alpha(t)))$ is substituted, 
the reference Hamiltonian $H_0(x,p,\alpha(t))$ 
is equal to the instantaneous energy as
\be
 H_0(x(t;\alpha(t)),p(t;\alpha(t)),\alpha(t))=E_0(\alpha(t)). \label{H0E0}
\ee
Then, by considering the time derivative, we find that 
the counterdiabatic term 
$H_{\rm CD}=H-H_0=\dot{\alpha}(t)\xi(x,p,\alpha)$, added to the Hamiltonian,
satisfies 
\be
 \frac{\partial H_0(x,p,\alpha)}{\partial\alpha}
 =\{\xi(x,p,\alpha),H_0(x,p,\alpha)\}+\frac{\diff E_0(\alpha)}{\diff\alpha},
 \label{di}
\ee
where $\{\cdot,\cdot\}$ denotes the Poisson bracket.
This is obtained by using the equation of motion (Section A of Ref.~\cite{sup}). 
If we find $\xi$ that satisfies this equation, 
we can realize an ideal time evolution that is characterized 
by $E_0(\alpha(t))$ at each time $t$.
This equation corresponds to the equation derived in Ref.~\cite{Jarzynski} 
and may be related to the equation 
for the dynamical invariant in the quantum STA~\cite{CRSCGM, LR}.
The commutator in the quantum system is replaced by the Poisson bracket 
in the classical limit.
We note that $\xi$ is not uniquely determined from Eq.~(\ref{di})~\cite{Jarzynski}.
This arbitrariness is discussed in the formulation discussed below.

It is a simple task to show that the adiabatic theorem holds in 
this counterdiabatic driving.
Taking the derivative of $J$ in Eq.~(\ref{J}) with respect to $\alpha$
and using Eq.~(\ref{di}), we obtain
\be
 \frac{\diff J}{\diff \alpha}
 = -\int \diff x\diff p\,
 \{\xi,H_0\}
 \delta\left(E_0-H_0\right).
\ee
This integral is evaluated by the surface contributions 
and goes to zero in systems with a smooth trajectory  
as we see in the following examples (Section A of Ref.~\cite{sup}).
This result means that $J$ is determined by the initial condition and
is independent of $t$.
The proof clearly indicates that  
the counterdiabatic term is introduced so that
the adiabatic theorem holds exactly.
We note that the time variable $t$ does not appear in Eq.~(\ref{di}) explicitly,
which allows us to handle the adiabatic invariant 
defined geometrically in phase space.
This result is the same as that in Ref.~\cite{Jarzynski}.

The solution of Eq.~(\ref{di}) can be studied systematically 
as was done in Ref.~\cite{OT} for the quantum system.
As an example, we set the reference Hamiltonian in a standard form 
\be
 H_0(x,p,\alpha(t))=p^2+U(x,\alpha(t)). \label{H0}
\ee
Then we show that the solution of Eq.~(\ref{di}) is given by the dispersionless 
Korteweg--de Vries (KdV) hierarchy~\cite{LM, Zhakharov}
(Section B of Ref.~\cite{sup}).
The corresponding method in the quantum STA 
was developed in Ref.~\cite{OT} and the KdV hierarchy was found.
The reference Hamiltonian and the counterdiabatic term 
represent the Lax pair in the corresponding nonlinear integrable system~\cite{Lax}.
The dispersionless KdV hierarchy is known 
as the ``classical'' limit of the KdV hierarchy~\cite{TT}.

When  $\xi(x,p,\alpha)$ is linear in $p$, the potential is of the form 
\be
 U(x,\alpha(t))=\frac{1}{\gamma^2(t)}u\left(\frac{x-x_0(t)}{\gamma(t)}\right),
 \label{sinvp}
\ee
where $u$ is an arbitrary function, 
and $\alpha$ represents both $x_0$ and $\gamma$, 
the former represents a translation and the latter a dilation.
The counterdiabatic term is given by
\be
 H_{\rm CD}=\dot{x}_0p+\frac{\dot{\gamma}}{\gamma}(x-x_0)p.
\ee 
This is known as the scale-invariant driving
and was found in previous works~\cite{delCampo, Jarzynski, DJdC}. 
A new result is obtained when we set that $\xi$ is third order in $p$.
We find that the counterdiabatic term is given by 
\be
 H_{\rm CD} =\dot{\alpha}\xi
 = \dot{\alpha}\left(pU(x,\alpha)+\frac{2}{3}p^3\right), \label{Hcddkdv}
\ee
and the potential satisfies the dispersionless KdV equation 
\be
 \frac{\partial U(x,\alpha)}{\partial \alpha}
 +U(x,\alpha)\frac{\partial U(x,\alpha)}{\partial x}=0. \label{dkdv}
\ee
This equation can be obtained by removing the third derivative term, 
the dispersion term, in the KdV equation~\cite{KdV}.
The form of the potential is different 
between the quantum and classical STA 
for the same form of the counterdiabatic term in Eq.~(\ref{Hcddkdv}).
This property is contrasted with that in the scale-invariant system where 
the quantum and classical STA give the same result.
In the same way, we can find the correspondence between 
the KdV and dispersionless KdV hierarchies at each odd order in $p$.
Although it is a difficult problem to implement the higher order terms 
in an actual experiment, 
some deformation of the counterdiabatic term is possible 
to represent the term by a potential function~\cite{OT}.


Before studying the solutions of the dispersionless KdV equation, 
we reformulate the classical STA by using the Hamilton--Jacobi theory.
The standard classical adiabatic theorem is described in periodic systems
since the validity of the approximation is written 
in terms of the period $T$ as $T|\dot{\alpha}/\alpha|\ll 1$. 
Although the adiabatic invariant is treated in ergodic 
systems~\cite{Hertz1, Hertz2, Ott}, 
its generalization is a delicate and difficult problem.
To find the quantum-classical correspondence of the adiabatic systems,
we need to extend the formulation to general systems.
This can be done by the Hamilton--Jacobi theory.
The adiabatic invariant is related to the action 
$S=\int_0^t \diff t'\,L =\int_0^t \diff t'\,(\dot{x}p-H)$.
This is a function of $x(t)$, $t$, and the whole history of $\alpha(t)$: 
$S=S(x(t),t,\{\alpha(t)\})$.
The property that $S$ is independent of the history of $x(t)$  
is shown by using the equation of motion.
In the counterdiabatic driving, 
trajectories in phase space are determined from Eq.~(\ref{H0E0}) and 
the Hamiltonian satisfies Eq.~(\ref{di}) that has no explicit time dependence.
These properties imply that the dynamics is characterized 
at each $t$, irrespective of past history.
By considering the variation $\alpha(t')\to \alpha(t')+\delta\alpha(t')$
of $S$ at an arbitrary $t'$ between 0 and $t$,  
and using the equation of motion, 
we obtain the deviation of the action as (Section C of Ref.~\cite{sup}) 
\be
 \delta S = \int_0^t \diff t'\, \delta \alpha(t')\left(
 -\frac{\partial H_0}{\partial\alpha}+\{\xi,H_0\} \right)
 -\left[\delta\alpha(t')\xi\right]_0^t. \no\\
\ee
We use Eq.~(\ref{di}) to find that the function defined as 
\be
 \Omega = S(x(t),t,\{\alpha(t)\})+\int_0^t\diff t'\,E_0(\alpha(t')) \label{Omega}
\ee
is independent of the history of $\alpha(t)$.
This function is a simple generalization 
of the Hamilton's characteristic function, or the abbreviated action, 
which is usually defined for constant $E_0$
by the Legendre transformation.
It satisfies 
\be
 \frac{\partial\Omega}{\partial x}=p(x,\alpha), \qquad
 \frac{\partial\Omega}{\partial \alpha}=-\xi(x,p(x,\alpha),\alpha).
\ee
The momentum $p$ is represented as a function of $x$ and $\alpha$
as we see from Eq.~(\ref{H0E0}).
These derivatives have no explicit $t$ dependence.
This implies that $\Omega$ is a function of $x$ and $\alpha$,
and not of $t$, just like the property of the Legendre transformation.
The explicit $t$ dependence of $\Omega$ can be removed by 
adding a time-dependent term, which does not change the trajectory, 
to the Hamiltonian.
As a result we can set $\Omega=\Omega(x(t),\alpha(t))$.

The Hamilton--Jacobi equation is given by 
$\frac{\partial S}{\partial t}+H=0$ with $p=\frac{\partial S}{\partial x}$.
We substitute $\Omega$ to this equation.
Noting that the time derivative is replaced in the present system with
$\partial_t\to \partial_t+\dot{\alpha}\partial_\alpha$, we find that 
the Hamilton--Jacobi equation for the counterdiabatic driving is decomposed as 
\be
 && H_0\left(x,p=\frac{\partial\Omega(x,\alpha)}{\partial x},\alpha\right)=E_0(\alpha), \label{HJ1}\\
 && \xi\left(x,p=\frac{\partial\Omega(x,\alpha)}{\partial x},\alpha\right)
 = -\frac{\partial\Omega(x,\alpha)}{\partial\alpha}.  \label{HJ2}
\ee
These equations are solved as a function of $x$ and $\alpha$, and have no
explicit time dependence.
We note that the counterdiabatic term is given by 
$H_{\rm CD}=\dot{\alpha}\xi(x,p,\alpha)$.
Thus the counterdiabatic driving is characterized by $\Omega$.

The definition of the action shows that $\Omega$ is written as 
\be
 \Omega(x(t),\alpha(t))
 &=&\int_0^t \diff t'\,\left(
 \dot{x}p(x,\alpha)-\dot{\alpha}\xi(x,p(x,\alpha),\alpha)\right) \no\\
 &=&\int_0^t \diff t'\,\left[
 \frac{\partial H_0}{\partial p}p
 +\dot{\alpha}\left(\frac{\partial \xi}{\partial p}p-\xi\right)\right],
 \label{omegaxa}
\ee
where we use the equation of motion in the second line.
This expression shows that
$\Omega$, as a function of $t$ and $\alpha(t)$, satisfies 
the following equations:
\be
 \left(\frac{\partial\Omega}{\partial t}\right)_\alpha 
 =  \frac{\partial H_0}{\partial p}p, \qquad
 \left(\frac{\partial\Omega}{\partial \alpha}\right)_t 
 =  \frac{\partial \xi}{\partial p}p-\xi. \label{xip}
\ee
Second equation states that $\Omega$ is independent of $\alpha(t)$
when the counterdiabatic term is linear in $p$.
This is the case of the scale-invariant driving where 
$H_0$ is of the form (\ref{H0}) with (\ref{sinvp}).
The Hamilton--Jacobi equation (\ref{HJ1}) reads 
\be
 \left(\frac{\partial\Omega}{\partial x}\right)^2
 +\frac{1}{\gamma^2}u\left(\frac{x-x_0}{\gamma}\right)
 =E_0(\alpha), 
\ee
and we find that $E_0$ and $\Omega$ take the form 
$E_0(\alpha(t))=\epsilon_0/\gamma^2(t)$ and 
$\Omega=\Omega((x-x_0)/\gamma)$, respectively.
By setting $\gamma(0)=1$, we can regard  
$\epsilon_0$ as the initial energy at $t=0$.
The counterdiabatic term is calculated as  
\be
 H_{\rm CD}=
 -\dot{x}_0\frac{\partial\Omega}{\partial x_0}
 -\dot{\gamma}\frac{\partial\Omega}{\partial \gamma}
 = \dot{x}_0p
 +\frac{\dot{\gamma}}{\gamma}(x-x_0)p, \label{cdsinv}
\ee
where we use the property that the derivatives of $\Omega$ 
with respect to $x_0$ and $\gamma$ 
are translated to that with $x$ in the present system.
$\Omega$ is also a function of $E_0(\alpha(0))=\epsilon_0$
and its definition shows that the derivative of $\Omega$ with respect to $\epsilon_0$ 
gives the relation 
\be
 \frac{\partial\Omega}{\partial\epsilon_0}
 =\int_0^t \frac{\diff t'}{\gamma^2(t')}
 =\tau(t),
\ee
where the last equality is the definition of the rescaled time $\tau(t)$.
We conclude that $\Omega$ as a function of $t$ and $\alpha$ 
in the scale-invariant system satisfies the relation 
\be
 \Omega(x(t;\alpha(t)),\alpha(t))
 =\Omega(x(\tau(t);\alpha(0)),\alpha(0)). \label{adinv}
\ee
The left-hand side represents $\Omega$ at $t$ 
obtained in the protocol $\alpha(t)$ 
and the right-hand side represents $\Omega$ at $\tau(t)$ 
in the fixed protocol $\alpha(0)$.
When the latter system gives a closed trajectory, 
$\Omega$ at the period is equal to the adiabatic invariant in Eq.~(\ref{J}) 
and is written as $\Omega=\oint p\diff x$ 
(Section D of Ref.~\cite{sup} for an example of the harmonic oscillator).
This relation shows that the adiabatic invariant is directly obtained from 
the corresponding nonperiodic trajectory.

For nonscale-invariant systems, Eq.~(\ref{adinv}) is not satisfied.
We treat the dispersionless KdV system as an example.
$H_0$ is given by Eq.~(\ref{H0}) 
and the potential $U$ satisfies the dispersionless KdV equation (\ref{dkdv}).
We can rederive Eq.~(\ref{Hcddkdv}) in the present formalism
by assuming that $E_0$ is constant.
Substituting $U=E_0-(\partial_x\Omega)^2$ to Eq.~(\ref{dkdv}), 
we find (Section C of Ref.~\cite{sup})
\be
 \frac{\partial\Omega}{\partial\alpha}
 +U\frac{\partial\Omega}{\partial x}
 +\frac{2}{3}\left(\frac{\partial\Omega}{\partial x}\right)^3 = 0. \label{omegadkdv}
\ee
This equation shows that the counterdiabatic term 
obtained from Eq.~(\ref{HJ2}) is given by Eq.~(\ref{Hcddkdv}).

Equation (\ref{dkdv}) is solved by the hodograph method as 
\be
 U(x,\alpha)=f\left(x-\alpha U(x,\alpha)\right),
\ee
where $f$ is an arbitrary function~\cite{Kodama, KG}.
As a simple example we consider the case $f(x)=x^2$.
Then, by solving the quadratic equation, we obtain 
\be
 U(x,\alpha)=\frac{2\alpha x+1-\sqrt{4\alpha x+1}}{2\alpha^2}. \label{quad}
\ee
We take the negative branch of the equation so that 
the trajectories are bound.
This potential is well-defined for $x>-\frac{1}{4\alpha}$ and we set 
the parameters so that this relation is satisfied throughout 
the time evolution.
By taking the limit $\alpha\to 0$, we have the harmonic oscillator 
$U(x,0)=x^2$.

\begin{center}
\begin{figure}[t]
\begin{center}
\includegraphics[width=1.\columnwidth]{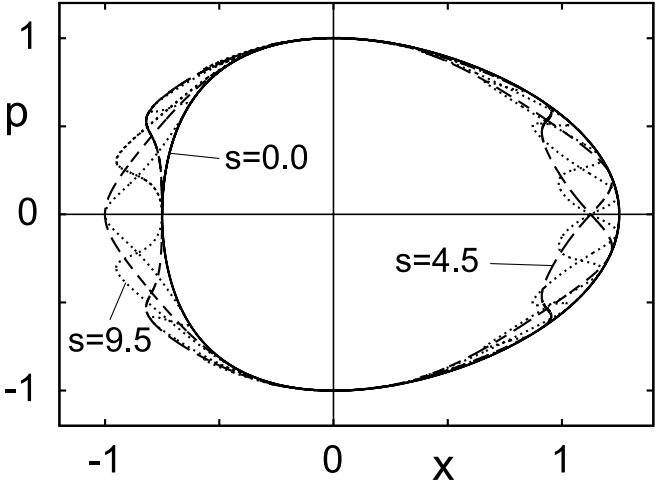}
\end{center}
\caption{Trajectories in phase space for the dispersionless KdV system 
in Eq.~(\ref{H0}) with (\ref{quad}).
We take the initial condition as $\alpha(0)=1/4$, 
$E_0(\alpha(0))=1$, and $(x(0),p(0))=(-3/4,0)$.
The solid line represents $s=0.0$, the dashed line $s=4.5$, 
and the dotted line $s=9.5$.}
\label{fig1:dkdv-phase}
\end{figure}
\end{center}
\begin{center}
\begin{figure}[t]
\begin{center}
\includegraphics[width=1.\columnwidth]{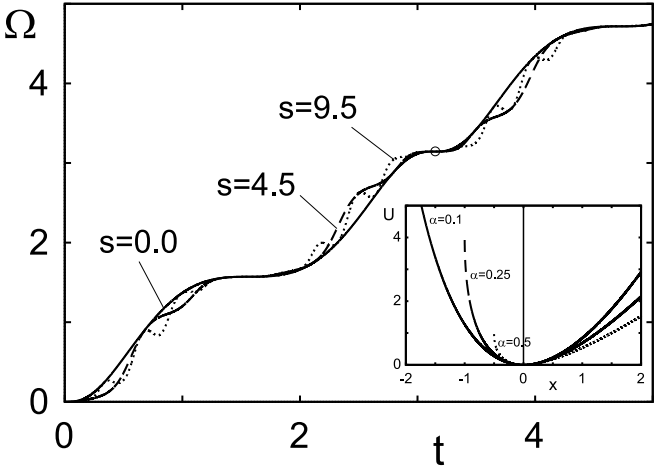}
\end{center}
\caption{The characteristic function $\Omega(x(t;\alpha(t)),\alpha(t))$
in the case of Fig.~\ref{fig1:dkdv-phase}.
The circle denotes the point $(T,\Omega(T))=(\pi,\pi)$ where 
$T$ denotes the period of the closed trajectory for a constant $\alpha$.
Inset: The potential function $U(x,\alpha)$ in Eq.~(\ref{quad}).}
\label{fig2:dkdv-omega}
\end{figure}
\end{center}

The equation of motion is solved numerically and we show
the trajectories in phase space and $\Omega(x(t;\alpha(t)),\alpha(t))$
in Figs.~\ref{fig1:dkdv-phase} and \ref{fig2:dkdv-omega}, respectively.
We use the protocol $\alpha(t)=\alpha(0)\left(1-\sin^2(st)\right)$.
We see from Fig.~\ref{fig2:dkdv-omega} that 
Eq.~(\ref{adinv}) is not satisfied since  
$\Omega$ is not necessarily a monotone increasing function 
and the time rescaling cannot give the result at $s=0$.
The counterdiabatic driving determines $\Omega$ instantaneously 
as we showed in the above analysis.
This can be confirmed numerically in the present system.
In Figs.~\ref{fig1:dkdv-phase} and \ref{fig2:dkdv-omega}, 
we consider an oscillating $\alpha(t)$ with the parameter $s$ and 
we see that $\Omega$ with $s$ is equal to $\Omega$ with $s=0$
when $t=2\pi n/s$ with integer $n$.
We also find that $\Omega$ is equal to the adiabatic invariant $J$ 
when $t$ is equal to the period of the closed trajectory which is defined 
for a fixed $\alpha$: 
\be
 \Omega(x(T;\alpha(T)),\alpha(T))=J.
\ee
This relation holds for an arbitrary choice of $\alpha(t)$ and 
can be proved by assuming that $T$ is independent of the initial energy.
We use the theory of action-angle variables for the proof.
See Sec.~E of Ref.~\cite{sup}. 
The period $T$ can also be calculated there and 
we find $T=\pi$ in the present case. 
Thus the adiabatic invariant can be calculated directly 
from the real nonperiodic trajectory.


The Hamilton--Jacobi theory of the classical STA 
makes a link between the classical and quantum systems.
In scale-invariant Hamiltonian in Eq.~(\ref{H0}) with (\ref{sinvp}), 
the counterdiabatic term in classical system 
becomes the same as that in quantum system 
if we use the symmetrization as $px\to (\hat{p}\hat{x}+\hat{x}\hat{p})/2$.
In the KdV systems, the form of the Hamiltonian is unchanged but 
the potential function $U$ 
satisfies an equation that is different from Eq.~(\ref{dkdv}).

In the quantum system the state is described by the wavefunction 
which satisfies the Schr\"odinger equation 
\be
 i\hbar\frac{\partial}{\partial t}\psi(x,t)
 = (\hat{H}_0+\hat{H}_{\rm CD})\psi(x,t).
\ee
For example, we consider the Hamiltonian (\ref{H0}) and (\ref{Hcddkdv})
with the replacement $pU\to(\hat{p}\hat{U}+\hat{U}\hat{p})/2$.
This setting gives a counterdiabatic driving 
when the potential satisfies the KdV equation~\cite{KdV}
\be
  \frac{\partial U(x,\alpha)}{\partial \alpha}
 +U(x,\alpha)\frac{\partial U(x,\alpha)}{\partial x}
 +\frac{\partial^3 U(x,\alpha)}{\partial x^3}
 =0.
\ee
By substituting the wavefunction 
\be
 \psi(x,t)= \e^{-iE_0t/\hbar}A(x,\alpha(t))\e^{i\Omega(x,\alpha(t))/\hbar},
\ee
where $A$ and $\Omega$ are real functions,
to the Schr\"odinger equation and taking the limit $\hbar\to 0$,
we can obtain Eq.~(\ref{omegadkdv}).
Thus the classical STA is obtained 
from the quantum version by taking the classical limit.
To find this relation, it is crucial to develop 
the classical STA by the Hamilton--Jacobi theory
as we discuss in this letter.

In the quantum STA, the wavefunction is given by the adiabatic state
of $H_0(\hat{x},\hat{p},\alpha(t))$.
By using the instantaneous eigenstate of $H_0$, $|n(\alpha(t))\rangle$,
and the instantaneous energy of $H_0$, $E_n(\alpha(t))$,
we can write the wave function as 
\be
 |\psi_n^{({\rm ad})}(t)\rangle
 = \exp\left(-\frac{i}{\hbar}\int_0^t\diff t'\,E_n(\alpha(t'))\right)
 |\tilde{\psi}_n(\alpha(t))\rangle, \no\\
\ee
 where
\be
 |\tilde{\psi}_n(\alpha(t))\rangle
 &=& \exp\left(-\int_{\alpha(0)}^{\alpha(t)} \diff \alpha'\,
 \langle n(\alpha')|\frac{\partial}{\partial\alpha'}|n(\alpha')\rangle\right)
 \no\\ && \times
 |n(\alpha(t))\rangle.
\ee
The point is that $|\tilde{\psi}_n(\alpha(t))\rangle$,
the adiabatic state without the dynamical phase, 
is written in terms of $\alpha$, not of $t$.
We define the unitary operator $\hat{V}$ as 
\be
 |\tilde{\psi}_n(\alpha(t))\rangle
 = \hat{V}(\alpha(t))|\tilde{\psi}_n(\alpha(0))\rangle.
\ee
This operator was introduced to develop the path integral formulation
of the adiabatic theorem~\cite{KNS}.
Then we can show that the counterdiabatic term is written as 
\be
 \hat{H}_{\rm CD}(t)=i\hbar\dot{\alpha}(t)
 \frac{\partial\hat{V}(\alpha)}{\partial \alpha}\hat{V}^\dag(\alpha).
\ee
We note that, in this expression, 
the unitary operator $\hat{V}$ can be 
replaced by the total time evolution operator
to find the formula by Demirplak and Rice~\cite{DR1}.
By using $\hat{V}$, we can write the formula of $\hat{H}_{\rm CD}$ 
in a more suggestive form.
We write $\hat{V}$ as 
\be
 \hat{V}(\alpha)=\exp\left(\frac{i}{\hbar}\hat{\Omega}(\alpha)\right), \label{V}
\ee
and this operator $\hat{\Omega}$ is the quantum analogue of the characteristic 
function $\Omega(x,\alpha)$ in the classical system.
For example, in the scale-invariant Hamiltonian
(\ref{H0}) with (\ref{sinvp}), $\hat{\Omega}$ is given by
\be
 && \hat{\Omega}(\alpha(t)) = -\hat{p}(x_0(t)-x_0(0)) \no\\
 && -\frac{1}{2}\left(\hat{p}(\hat{x}-x_0(t))+(\hat{x}-x_0(t))\hat{p}\right)
 \ln\frac{\gamma(t)}{\gamma(0)}.
\ee
The counterdiabatic term in this case is written as 
\be
 \hat{H}_{\rm CD}(t)=-\dot{\alpha}(t)
 \frac{\partial\hat{\Omega}(\alpha)}{\partial\alpha}.
 \label{s0hat}
\ee
This expression formally coincides with Eq.~(\ref{HJ2}).


To summarize, we have developed the classical STA 
by using the Hamilton--Jacobi theory.
The system is characterized by 
the generalized characteristic function $\Omega(x,\alpha)$
and the counterdiabatic term is obtained from this function.
The equation for the counterdiabatic term can be studied systematically 
and is solvable when the system falls in the dispersionless KdV hierarchy.
Our formulation also gives a relation to the adiabatic theorem.
We can also show that the classical STA is reduced from 
the quantum STA by taking the standard semiclassical approximation.

We acknowledge financial support from the 
ImPACT Program of the Council for Science, 
Technology, and Innovation, Cabinet Office, Government of Japan.
K.T. was supported by JSPS KAKENHI Grant No. 26400385.


\onecolumngrid
\newpage

\makeatletter
\renewcommand{\thefigure}{\thesection.\arabic{figure}}
\makeatother

\begin{center}
{\bf \large Supplemental Material for \\ 
``Quantum-Classical Correspondence of Shortcuts to Adiabaticity''} 
\end{center}
\begin{center}
Manaka Okuyama and Kazutaka Takahashi
\end{center}

\bigskip
\bigskip

\twocolumngrid
\section{A. Classical STA}
\setcounter{equation}{0}
\def\theequation{A.\arabic{equation}}

We start from Eq.~(2) to derive the formula of the classical STA.
Taking the time derivative, we have 
\be
 \frac{\diff H_0(x,p,\alpha)}{\diff t}
 =\dot{\alpha}(t)\frac{\diff E_0(\alpha)}{\diff\alpha}.
\ee
The left-hand side is rewritten by using the equations of motion 
$\dot{x}=\frac{\partial H}{\partial p}$ and 
$\dot{p}=-\frac{\partial H}{\partial x}$ as 
\be
 \dot{\alpha}\frac{\partial H_0}{\partial\alpha}
 -\left\{H,H_0\right\}
 =\dot{\alpha}\frac{\diff E_0}{\diff\alpha}. 
\ee
This gives Eq.~(3).

The adiabatic theorem is proved as follows.
We take the derivative with respect to $\alpha$ of Eq.~(1) to find 
\be
 \frac{\diff J}{\diff \alpha}
 = \int \diff x\diff p\,
 \left(\frac{\diff E_0}{\diff\alpha}
 -\frac{\partial H_0}{\partial\alpha}\right)
 \delta\left(E_0-H_0\right).
\ee
Using Eq.~(3), we can write 
\be
 \frac{\diff J}{\diff \alpha}
 &=& -\int \diff x\diff p\,
 \{\xi,H_0\}
 \delta\left(E_0-H_0\right) \no\\
 &=& \int \diff x\diff p\,
 \left(
 \frac{\partial\xi}{\partial x}
 \frac{\partial}{\partial p}
 -\frac{\partial\xi}{\partial p}
 \frac{\partial}{\partial x}\right)
 \theta\left(E_0-H_0\right). \no\\
\ee
This integration is written by the surface contributions as 
\be 
 \frac{\diff J}{\diff \alpha}
 &=& \int \diff x\,\left[\frac{\partial\xi}{\partial x}
 \theta\left(E_0-H_0\right) \right]_{p_-}^{p_+}\no\\
 && -\int \diff p\,\left[\frac{\partial\xi}{\partial p}
 \theta\left(E_0-H_0\right) \right]_{x_-}^{x_+}.
\ee
When $x$ is equal to its maximum or minimum value, $x_+$ or $x_-$, 
$p$ takes a fixed value for  a smooth trajectory.
In this case, this derivative goes to zero and we can prove the adiabatic theorem.

\section{B. Dispersionless KdV hierarchy}
\setcounter{equation}{0}
\def\theequation{B.\arabic{equation}}

For the reference Hamiltonian $H_0$ in Eq.~(5), 
we study possible forms of the potential $U$ and the corresponding 
counterdiabatic term $H_{\rm CD}=\dot{\alpha}\xi(x,p,\alpha)$.
$\xi$ is obtained by solving Eq.~(3).
In the present case, it is written as 
\be
 \left(-\frac{p}{m}\frac{\partial}{\partial x}
 +\frac{\partial U}{\partial x}
 \frac{\partial}{\partial p}\right)\xi
 = 
 \frac{\diff E_0}{\diff\alpha} 
 -\frac{\partial U}{\partial \alpha}. \label{di2}
\ee
This equation shows that $\xi$ is odd order in $p$.
We note that the additional relation $p^2 = E_0(\alpha)-U(x,\alpha)$ 
is used to solve this equation.

As an example, we first assume that $\xi$ is linear in $p$: 
\be
 \xi(x,p,\alpha)=p\tilde{\xi}_0(x,\alpha).
\ee
We also assume 
\be
 && E_0(\alpha)=\epsilon_0 \alpha^k, 
\ee
where $\epsilon_0$ is constant.
Equation (\ref{di2}) reads 
\be
 \left[-2(E_0-U)\frac{\partial}{\partial x}
 +\frac{\partial U}{\partial x}
 \right]\tilde{\xi}_0
 =-\frac{\partial U}{\partial\alpha}
 +\frac{k}{\alpha}E_0.
\ee
This equation holds for an arbitrary choice of initial energy 
and we obtain 
\be
 \tilde{\xi}_0(x,\alpha)=-\frac{k}{2\alpha}x+c_0(\alpha),
\ee
where $c_0$ is an arbitrary function of $\alpha$.
The potential function satisfies 
\be
 \alpha\frac{\partial U}{\partial\alpha}
 +\left(-\frac{k}{2}x+\alpha c_0(\alpha)\right)
 \frac{\partial U}{\partial x} -kU =0.
\ee
Substituting $U(x,\alpha)=\alpha^k \tilde{U}(x,\alpha)$ to 
this equation, we obtain 
\be
 \alpha\frac{\partial \tilde{U}}{\partial\alpha}
 +\left(-\frac{k}{2}x+\alpha c_0(\alpha)\right)
 \frac{\partial \tilde{U}}{\partial x} =0.
\ee
This equation shows that the potential has the scale-invariant form 
\be
 U=\alpha^k\tilde{U}=\alpha^k U_0\left(\frac{x-x_0(\alpha)}{\alpha^{k/2}}\right),
\ee
where $U_0$ is an arbitrary function and $x_0$ is determined from the equation 
\be
 \alpha^k \frac{\diff x_0(\alpha)}{\diff \alpha}-\frac{k}{2}x_0(\alpha)
 = \alpha c_0(\alpha).
\ee
By using a proper reparametrization of the parameters, we obtain 
the potential in Eq.~(6).

We next consider the case where $\xi$ is third order in $p$: 
\be
 && \xi(x,p,\alpha)=p\tilde{\xi}_0(x,\alpha)+p^3\tilde{\xi}_2(x,\alpha).
\ee
We also assume that $E_0$ is constant: 
\be
 \frac{\diff E_0(\alpha)}{\diff \alpha}=0.
\ee
Then Eq.~(3) is written as 
\be
 \left(-2p^2\frac{\partial}{\partial x}
 +\frac{\partial U}{\partial x}\right)\tilde{\xi}_0
 +\left(-2p^4\frac{\partial}{\partial x}
 +3p^2\frac{\partial U}{\partial x}
 \right)\tilde{\xi}_2
 = 
 -\frac{\partial U}{\partial \alpha}, \no\\
\ee
and we obtain the conditions 
\be
 &&\frac{\partial \tilde{\xi}_2}{\partial x}
 = 0, \\
 &&
 -\frac{1}{m}\frac{\partial\tilde{\xi}_0}{\partial x}
 +3\frac{\partial U}{\partial x}\tilde{\xi}_2
 = 0, \\
 &&
 \frac{\partial U}{\partial x}\tilde{\xi}_0
 = 
 -\frac{\partial U}{\partial \alpha}.
\ee
The first equation shows that $\tilde{\xi}_2$ is independent of $x$: 
\be
 \tilde{\xi}_2(x,\alpha)=c_2(\alpha), 
\ee
the second equation shows that $\tilde{\xi}_2$ is written in terms of $U$: 
\be
 \tilde{\xi}_0=3m c_2(\alpha)U+c_0(\alpha), 
\ee
and the third equation gives  
\be
 \frac{1}{3mc_2(\alpha)}\frac{\partial U}{\partial \alpha} 
 +\left(U+\frac{c_0(\alpha)}{3mc_2(\alpha)}\right)
 \frac{\partial U}{\partial x}
 = 0.
\ee 
This is essentially equivalent to the dispersionless KdV equation in Eq.~(9).

We can go further to derive the higher order 
dispersionless KdV equations.
They are fifth order  in $p$, seventh order, and so on.

\section{C. Hamilton-Jacobi theory}
\setcounter{equation}{0}
\def\theequation{C.\arabic{equation}}

We briefly review the Hamilton-Jacobi theory to formulate the classical STA.
The system is characterized by the action defined from the Lagrangian: 
\be
 S=\int_0^t \diff t'\,L(x(t'),\dot{x}(t');\alpha(t)).
\ee
It satisfies, by definition,  
\be
 \frac{\diff S}{\diff t}=L. \label{eq1}
\ee
On the other hand we can write 
\be
 \frac{\diff S}{\diff t}
 =\frac{\partial S}{\partial x}\dot{x}
 +\frac{\partial S}{\partial t}
 =\frac{\partial L}{\partial \dot{x}}\dot{x}
 +\frac{\partial S}{\partial t}, \label{eq2}
\ee
where we use the conjugate momentum 
\be
 p = \frac{\partial L}{\partial \dot{x}}= \frac{\partial S}{\partial x}.
 \label{p}
\ee
The first equality is the definition of the momentum and the second 
is derived from the variation of the action, $x(t')\to x(t')+\delta x(t')$ as 
\be
 \delta S &=& \int \diff t'\, \left(
 \delta x(t')\frac{\partial L }{\partial x}
 +\delta \dot{x}(t')\frac{\partial L }{\partial \dot{x}}\right) \no\\
 &=& \int \diff t'\, \delta x(t')\left(
 \frac{\partial L }{\partial x}
 -\frac{\diff}{\diff t'}\frac{\partial L }{\partial \dot{x}}\right)
 +\left[\delta x(t')\frac{\partial L}{\partial\dot{x}}\right]_0^t. \no\\
\ee
The first term goes to zero if we use the equation of motion.
This means that the action is independent of the history of $x$ and
is determined as a function of $x(t)$.
The derivative of the action with respect to $x(t)$ gives the second 
equality in Eq.~(\ref{p}).

In a similar way we can consider the variation 
$\alpha(t')\to \alpha(t')+\delta \alpha(t')$: 
\be
 \delta S 
 &=& \int_0^t \diff t'\,
 \delta\alpha\left[
 \frac{\partial p}{\partial\alpha}
 \left(\dot{x}-\frac{\partial H}{\partial p}\right)
 -\frac{\partial H_0}{\partial \alpha}
 -\dot{\alpha}\frac{\partial \xi}{\partial \alpha} \right.\no\\
 && \left.
 +
 \dot{x}\frac{\partial\xi}{\partial x}
 +\dot{p}\frac{\partial\xi}{\partial p}
 +\dot{\alpha}\frac{\partial\xi}{\partial \alpha}
 \right]
 -\left[\delta \alpha\xi\right]_0^t.
\ee
Using the equation of motion and
the condition for the counterdiabatic term in Eq.~(3),
we obtain  
\be
 \delta S 
 &=& \int_0^t \diff t'\,\delta \alpha\left(
 -\frac{\partial H_0}{\partial \alpha}
 +\{\xi,H_0\}
 \right)
 -\left[\delta \alpha\xi\right]_0^t \no\\
 &=& -\int_0^t \diff t'\,\delta \alpha
 \frac{\diff E_0(\alpha)}{\diff\alpha}  
 -\left[\delta \alpha\xi\right]_0^t.
\ee
This shows that $\Omega$ defined as Eq.~(11) is independent of 
the history of $\alpha(t)$.

The Hamilton-Jacobi equation is derived 
by equating Eqs.~(\ref{eq1}) and (\ref{eq2}).
Using the definition of the Hamiltonian $H=\dot{x}p-L$, we obtain 
\be
 \frac{\partial S(x,t,\{\alpha(t)\})}{\partial t}
 +H\left(x,p=\frac{\partial S}{\partial x};\alpha(t)\right) = 0.
\ee
We note that the action is dependent on $x$, $t$, and $\alpha(t)$, and 
the time derivative acts on $\alpha$.
By replacing $\partial_t$ with $\partial_t+\dot{\alpha}\partial_\alpha$, 
we write 
\be
 &&\frac{\partial S(x,t,\{\alpha(t)\})}{\partial t}
 +H_0\left(x,p=\frac{\partial S}{\partial x},\alpha(t)\right)  \no\\
 &&+\dot{\alpha}\left(
 \frac{\partial S(x,t,\{\alpha(t)\})}{\partial \alpha}
 +\xi\left(x,p=\frac{\partial S}{\partial x},\alpha(t)\right)\right) = 0. \no\\
\ee
This equation and the definition of $\Omega$ show that 
the Hamilton-Jacobi equation is decomposed 
as Eqs.~(13) and (14).

The derivation of the counterdiabatic term of the scale-invariant system
was done in Eq.~(18).
Here we consider the case of the dispersionless KdV equation.
Substituting $U=E_0-(\partial_x\Omega)^2$ to Eq.~(9)
and assuming $E_0$ is constant, we have
\be
 -2\frac{\partial\Omega}{\partial x}
 \left[
 \frac{\partial}{\partial x}\left(
 \frac{\partial\Omega}{\partial\alpha}+U\frac{\partial\Omega}{\partial x}\right)
 +2\frac{\partial^2\Omega}{\partial x^2}\frac{\partial \Omega}{\partial x}
 \right]=0. \no\\
\ee
This gives 
\be
 \frac{\partial}{\partial x}\left[
 \frac{\partial\Omega}{\partial\alpha}+U\frac{\partial\Omega}{\partial x}
 +\frac{2}{3}\left(\frac{\partial \Omega}{\partial x}\right)^3
 \right]=0. 
\ee
By gauging out an irrelevant constant term, we obtain Eq.~(21).

\section{D. Harmonic oscillator}
\setcounter{equation}{0}
\def\theequation{D.\arabic{equation}}

\setcounter{figure}{0}
\def\thefigure{D.\arabic{figure}}
\begin{center}
\begin{figure}[t]
\begin{center}
\includegraphics[width=1.\columnwidth]{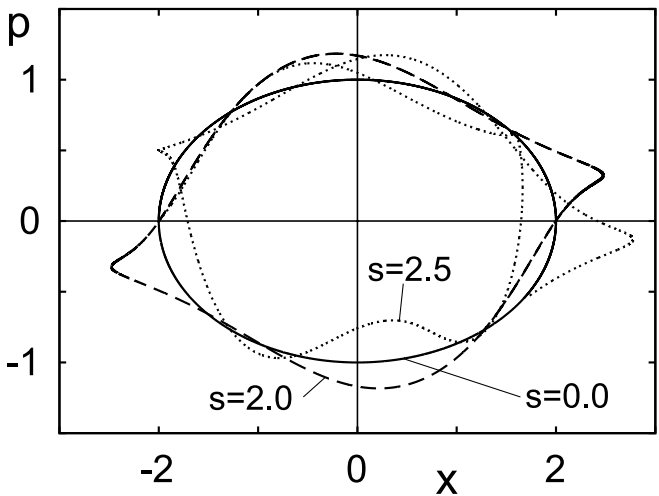}
\end{center}
\caption{Trajectories in phase space for the harmonic oscillator 
in Eq.~(\ref{ho}).
We use the protocol $\omega(t)=1+\frac{1}{2}\sin (st)$, 
and take the initial condition as 
$E_0(\omega(0))=1$ and $(x(0),p(0))=(-2,0)$.}
\label{fig1:ho-phase}
\end{figure}
\end{center}
\begin{center}
\begin{figure}[t]
\begin{center}
\includegraphics[width=1.\columnwidth]{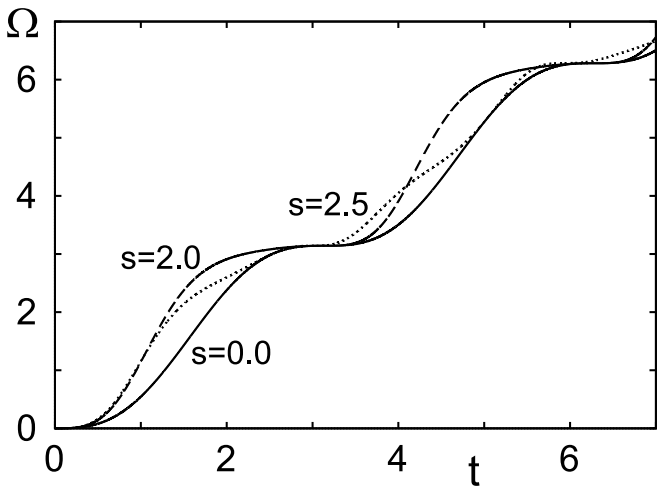}
\end{center}
\caption{The characteristic function $\Omega(x(t;\omega(t)),\omega(t)$
in the case of Fig.~\ref{fig1:ho-phase}.}
\label{fig2:ho-omega}
\end{figure}
\end{center}

We consider an example of the harmonic oscillator
\be
 H_0=p^2+\frac{\omega^2(t)}{4}x^2, \label{ho}
\ee
with $\omega(t)=1+\frac{1}{2}\sin (st)$.
This system is described as a scale-invariant system and the parameter 
$\omega$ represents the dilation effect as $\gamma=1/\sqrt{\omega}$.
At $s=0$, we have a static system and the trajectory in phase space 
becomes a closed curve.
This is not the case for systems with $s\ne 0$ 
as we see in Fig.~\ref{fig1:ho-phase}.
$\Omega(t)$ for several values of $s$ are plotted 
in Fig.~\ref{fig2:ho-omega}.
By changing the horizontal axis as $\tau(t)=\int_0^t \diff t'\,\omega(t')$, 
we can confirm that all curves fall on the curve at $s=0$.
In the present example, 
$\Omega$ can be calculated analytically and is given by 
\be
 \Omega(x(t;\omega(t)),\omega(t))
 =\frac{E_0(\omega(t))}{\omega(t)}\left(\tau(t)-\sin\tau(t)\cos\tau(t)\right).
 \no\\
\ee
We can show Eq.~(20) by using the relation
$\frac{E_0(\omega(t))}{\omega(t)}=\frac{E_0(\omega(0))}{\omega(0)}$.

\section{E. Action-angle variables}
\setcounter{equation}{0}
\def\theequation{E.\arabic{equation}}

The adiabatic invariant $J$ 
is independent of the parameter $\alpha(t)$.
It is a function of the initial energy $E_0(\alpha(0))=\epsilon_0$: 
$J=J(\epsilon_0)$.
Then the characteristic function is written as a function of $x$, $\alpha$, and $J$: 
$\Omega = \Omega(x,\alpha,J)$.

In the periodic systems, we know that 
the action-angle variables are useful to characterize the system.
We define the angle variable 
\be
 w = \frac{\partial\Omega}{\partial J}.
\ee
In the theory of canonical transformation, $w$ and the action variable $J$ 
are interpreted as the canonical variables.
$J$ represents the conjugate momentum of $w$.

We assume, for simplicity,  
that the instantaneous energy is constant: $E_0(\alpha(t))=\epsilon_0$.
In this case, the angle variable is represented as 
\be
 w = \frac{\partial\epsilon_0}{\partial J}\frac{\partial\Omega}{\partial \epsilon_0}
 = \frac{\partial\epsilon_0}{\partial J}t. \label{w}
\ee
We note that $J(\epsilon_0)$ is independent of $\alpha(t)$.
This means that $w$ is proportional to $t$: 
$w = t/T(J)$.
Taking the derivative of $w$ with respect to $x$, we obtain  
\be
 \frac{\partial w}{\partial x}
 = \frac{\partial^2\Omega}{\partial x\partial J}
 = \frac{\partial p}{\partial J}.
\ee
For a fixed $\alpha$, we consider the integration of $w$ over 
the closed trajectory.
Then we obtain 
\be
 w = \oint \diff x\,\frac{\partial w}{\partial x}
 = \oint \diff x\,\frac{\partial p}{\partial J}
 = \frac{\partial}{\partial J}\oint \diff x\,p
 = 1.
\ee
This result shows that $T$ is equal to the period of the trajectory.

Integrating $w$ with respect to $J$, we write 
\be
 \Omega(x,\alpha,J)=\int_{0}^J \diff J'\,\frac{t(x,\alpha,J')}{T(J')},
 \label{OmegaJ}
\ee
where we set the boundary condition $\Omega=0$ at $J=0$.
We note that this expression does not mean that $\Omega$ is proportional to $t$.
$t$ in the right hand side is defined as  
$t=\frac{\partial\Omega}{\partial\epsilon_0}$ and is a function of 
$x$, $\alpha$, and $J$.
Thus Eq.~(\ref{OmegaJ}) has no explicit time dependence and 
the real dynamics is not implemented in this expression.

Now we substitute $x$ and $\alpha$ at $t=T(J)$ to Eq.~(\ref{OmegaJ}).
Then 
\be
 && \Omega(x(T(J),J),\alpha(T(J)),J) \no\\
 &=& \int_{0}^J \diff J'\,\frac{t(x(T(J),J),\alpha(T(J)),J')}{T(J')}.
\ee
We note that $x$ is a function of $t$ and $J$.
If $T$ is independent of $J$, the function $t$ in the right-hand side of
this equation must be equal to $T$.
We obtain in that case 
\be
 \Omega(x(T,J),\alpha(T),J) = J. 
\ee

As an example, we consider the dispersionless KdV system with the potential 
in Eq.~(23).
The adiabatic invariant is represented as 
\be
 J=2\int_{x_-}^{x_+} \diff x\,\sqrt{\epsilon_0-U(x,\alpha)},
\ee
where $x_\pm$ represent the end points of the trajectory.
Since the potential has the form $U(x,\alpha)=u(\alpha x)/\alpha^2$, we can write 
\be
 J = \frac{2}{\alpha^2}\int_{z_-}^{z_+} 
 \diff z\,\sqrt{\epsilon_0\alpha^2-u(z)},
\ee
where $z_\pm =\alpha x_\pm$.
We know from the adiabatic theorem that this function is independent of $\alpha$.
Then we can evaluate this integral by taking the limit $\alpha\to 0$.
$u(z)$ is replaced with $z^2$ and we find 
\be
 J=\frac{2}{\alpha^2}\frac{\pi\epsilon_0\alpha^2}{2}
 =\pi\epsilon_0.
\ee
This shows that the period of the trajectory is given by 
\be
 T=\frac{\partial J}{\partial\epsilon_0}=\pi.
\ee
This is independent of the initial energy $\epsilon_0$.

\end{document}